\documentclass[article,twocolumn,amsmath,floats,showpacs,nofootinbib]{revtex4}
\usepackage[usenames]{color}

\usepackage{graphicx}
\usepackage{epstopdf}
\usepackage{textcomp}
\usepackage{hyperref}
\usepackage{multirow}

\hypersetup{colorlinks=true}

\begin{document}

\title{Point-contact spectra of $2H-NbS{{e}_{2}}$ in the superconducting state}

\author{N. L. Bobrov, L. F. Rybal'chenko, M. A. Obolenskii, and V. V. Fisun}
\affiliation{B.I.~Verkin Institute for Low Temperature Physics and
Engineering, of the National Academy of Sciences
of Ukraine, prospekt Lenina, 47, Kharkov 61103, Ukraine and A. M. Gor'kov State University, Khar'kov
Email address: bobrov@ilt.kharkov.ua}
\published {(\href{http://fntr.ilt.kharkov.ua/fnt/pdf/11/11-9/f11-0925r.pdf}{Fiz. Nizk. Temp.}, \textbf{11}, 897 (1985)); (Sov. J. Low Temp. Phys., \textbf{11}, 510 (1985)}
\date{\today}

\begin{abstract}A qualitative study of the electron-phonon interaction (EPI) spectra in $2H-NbS{{e}_{2}}$  is carried out by the method of point-contact spectroscopy in the superconducting state.
Anisotropy of the EPI spectra for contacts oriented in the principal crystallographic directions (along the $c$ axis and parallel to the basal surface) was observed. Approximate values of the positions
of the main features in the phonon density of states were established for both directions. The additional low-frequency peaks observed in the spectra for separate crystallographic orientations are
associated with the collective oscillations of the amplitude and phase of charge density waves.

\end{abstract}

\maketitle

\section{INTRODUCTION}
In the solution of problems associated with high-temperature superconductivity, until recently, great significance was ascribed to low-dimensional materials. It was believed that the exciton mechanism
of superconductivity could be realized in quasi-two-dimensional crystals with alternating layers of atoms of the metal and of the semiconductor \cite{Little,Ginzburg}. At the same time, not enough attention was devoted
to the study of the characteristics of the electron-phonon interaction (EPI) in these materials. Thus for $2H-NbS{{e}_{2}}$   only the dispersion curves $\omega (q)$  were obtained by the method of
slow-neutron diffraction for several tranches of the lattice vibrations in a restricted range of wave numbers \cite{Moncton}. The position of some points on the dispersion curves was also established in
experiments on Raman light scattering \cite{Tsang}. A calculation of the dispersion curves in the normal state (above the Peierls transition temperature) was carried out in two studies for the $2H-NbS{{e}_{2}}$
lattice \cite{Feldman,Motizuki}. In the first of these studies, using a purely phenomenological approach, $\omega (q)$ for the $\Delta $  and $\Sigma $  directions (along the normal to the layers and perpendicular to the
lateral face of the Brillouin zone in the basal surface, respectively) were calculated. In the second study, \cite{Motizuki} $\omega (q)$ along the $\Delta $, $\Sigma $, and $T$ axes (the latter is oriented from
the center to the vertex of the Brillouin zone in the basal surface) was calculated in the rigid-ion model, though for separate vibrational branches (associated with the lattice distortions
accompanying the Peierls transition) the interaction of conduction electrons with the lattice was also taken into account. At the same time, just as in the first study, a number of the parameters of
the dynamic matrix were chosen from the experimental data on light scattering and scattering of slow neutrons in the low-frequency range. This is apparently one of the reasons that the results of the
calculations of $\omega (q)$ in both studies are not in satisfactory agreement in the high-frequency region.

A peculiarity of the phonon spectrum of dichalcogenides of transition metals, including $2H-NbS{{e}_{2}}$, is the presence of a "soft" mode in the $\Sigma $  direction at comparatively low temperatures
- a weak trough in the dependence $\omega (q)$ near $q={2}/{3{{q}_{\max }}}\;$ \cite{Moncton}. As the temperature is lowered, further softening of this mode, terminating in a structural phase transition of the
Peierls type, occurs. (For $2H-NbS{{e}_{2}}$ the transition temperature is equal to $\sim 33 K$.) At the same time a triplet chargedensity wave (CDW) in the spatial distribution of $d-$ electrons forms on
the $\Sigma $ direction, and corresponding static distortions of the lattice appear: a superstructure with a period approximately equal to three times the period of the starting lattice (the so-called
incommensurateness) is equal to several percent and for $NbS{{e}_{2}}$ it remains down to $1.3\text{ }K$ (Ref. \cite{Bulaevskii}). The appearance of a superlattice gives rise to new vibrational oranches in the $\Sigma
$ direction which, however, was not taken into account in the calculations of $\omega (q)$ \cite{Feldman,Motizuki}.

The method of point-contact spectroscopy is widely used to study the EPI in normal metals \cite{Yanson1}. It has turned out \cite{Yanson2,Yanson3} that analogous studies can also be carried out for metals in the superconducting
state. In this case, practically the same results were obtained for pure contacts of the $S-c-N$ type ($S$ is the superconductor, $c$ is the constriction, and $N$ is the normal metal). In the case of
$S-c-S$ contacts the phonon features (without the parasitic peaks) have been observed only in the dirty limit, when superconductivity in the region of the constriction was substantially suppressed \cite{Yanson3}.
In this case the phonon structure in the point-contact spectra (${{V}_{2}}\left( eV \right)\sim {{d}^{2}}V/d{{I}^{2}}$) is substantially amplified compared with the $N$ state, though the form of the
peaks changes appreciably. Approximately the same situation also occurs in the case of dirty $S-c-N$ contacts.

The purpose of this work is to establish the form of the point-contact spectra in $2H-NbS{{e}_{2}}$ superconducting point contacts in two basic crystallographic directions (perpendicular and parallel
to the basal surfaces) and to compare the phonon features appearing in the spectra with the published data available for them.

The studies performed established the positions of the main phonon peaks for the principal symmetry axes of the crystal lattice of $NbS{{e}_{2}}$ on the energy axis. Low-temperature peaks, which we
associated with the collective oscillations of the amplitude and phase of the CDW in the low-temperature modification of $NbS{{e}_{2}}$, were observed for a number of spectra for one of the
crystallographic orientations.
\section{EXPERIMENTAL PROCEDURE}

The I-V curves and their second derivatives (the point-contact spectra) of two types of point contacts were studied: $NbS{{e}_{2}}-NbS{{e}_{2}}$ and $NbS{{e}_{2}}-Cu$. The use of $Cu$ as one of the electrodes enabled lowering to a significant extent the thermal effects in the contact region. At the same time it was possible to measure the point-contact characteristics in both basic crystallographic directions of $NbS{{e}_{2}}$: along the layers and perpendicular to them (i.e., in a direction coinciding with the $c$ axis). We note that the use of contacts of only the first type would methodically exclude the possibility of measuring spectra along the $c$ axis.

The starting $NbS{{e}_{2}}$ consisted of a $20\text{ }\times 15\text{ }\times \text{ }0.1\text{ }mm^3$ single-crystalline plate. From it we cut out $5\times 3\,mm^2$  electrodes with a knife and, using silver paste, we glued them to a wire $\Gamma $-shaped damper made of beryllium bronze $\sim 0.2-0.3\text{ }mm$ in diameter with each arm $\sim 15\text{ }mm$ long. Thus the contact areas of the electrodes either had natural crystal faceting or were worked mechanically. For the measurements along the $c$ axis the copper electrode consisted of a three-facet prism with a $2-mm$ base and $4-mm$ high, and for measurements along the layers it consisted of a $2\times 2\times 10-mm^3$ parallelepiped, cut out of a single crystal by the electric-spark method. The copper electrode was electrolytically polished in a mixture of concentrated nitric and phosphoric acids and acetic anhydride, taken in the volume ratio 2:1:1, with a voltage of 10-15 $V$. The electrode was then washed in distilled water, and the quality of its surface was monitored under a microscope with 100-fold magnification in oblique light. The working surface should not have traces of contamination in this case. The radius of the tip of the prism after the electrolytic polishing was equal to 0.1-0.2 $mm$.

After the copper electrode was prepared, the samples were immediately mounted into a clamping device. For the measurements along the c axis the tip of the three- facet copper prism was oriented perpendicular to the flat surface of the $NbS{{e}_{2}}$ electrode, and a more rigid damper was used. For measurements along the layers the contacts were formed between the edge of the copper parallelepiped and the face of the plate or between two faces of the flat $NbS{{e}_{2}}$ samples. We used a softer damper to prevent crushing of the layers. It should be emphasized that the electrochemical treatment of the copper electrode must be carried out prior to each measurement, since even in a cryostat in a helium atmosphere a much too thick layer of oxide forms on the surface of the electrode, and in order to create a point contact a large force must be applied to clamp electrodes to one another, which crushes the $NbS{{e}_{2}}$ layers.

The point contacts were created by the shear method \cite{Yanson1}, and the clamping force in this case was regulated outside the cryostat with the help of a torsion setup, which permitted twisting one of the arms of the $\Gamma $-shaped diameter.

We performed the measurements in the absence of a magnetic field in one of two temperature intervals: 1.6-4.2 or $4.4-10\,K$. The first temperature interval was obtained by submerging the contacts into liquid helium followed by pumping out of the vapor and in the second case we used an intermediate cryostat with gas heat exchange. To measure the point-contact characteristics in the $N$ state the temperature was raised above the temperature of the superconducting transition in $NbS{{e}_{2}}$.

The second derivatives of the I-V curves were measured using a standard modulation technique and were recorded on a X-Y automatic plotter.
\section{EXPERIMENTAL RESULTS}
\subsection{Peculiarities of point-contact spectroscopy of $2H-NbS{{e}_{2}}$ }
It was previously established \cite{Yanson3} that the main condition for carrying out the point-contact spectroscopy of EPI in the superconducting state is that both the heating effect and the effects caused by the jump-like destruction of superconductivity in the contact region when the electric bias is increased must be small. The latter effects are manifested in the form of a jump-like decrease in the excess current for some values of the voltage. At the same time the true phonon peaks can be mashed by much more intense (we shall call them "critical") features. One reason that effects of the second type appear could be the nonuniform structure of the region of the constriction both of a geometrical and chemical character, since the corresponding features were observed even in quite clean contacts \cite{Khotkevich}.

It turned out that the "critical" features are virtually absent in contacts in which superconductivity is substantially suppressed in the region of the constriction and thermal effects are insignificant. Such a compromise can be achieved with the appearance of effective centers of depairing in the region of the constriction. In the case of $Nb$ and, most likely, compounds based on this element, strong breaking of Cooper pairs can occur also at the boundary between the nonstoichiometric niobium oxide and the superconductor as a result of the formation of surface states of the $d$ type localized there \cite{Halbritter}.

Measurements on point contacts made of $NbS{{e}_{2}}$ in the superconducting state ($S-c-S$ contacts) showed that in spite of the comparatively low purity of the starting material (${{\rho }_{300}}/{{\rho }_{\text{res}}}\sim 30$), the ballistic current flow in such contacts can be realized without special difficulties. These latter were as a rule, accompanied by strong "critical" effects. The creation of such (undesirable in this case) pure contacts was facilitated by the use of fresh chips or sections of $NbS{{e}_{2}}$  plates for the contact surfaces. For this reason, in order to obtain dirty contacts (we have in mind contacts with the superconductivity substantially suppressed in the region of the constriction), after mechanical preparation of the surface under study, we oxidized the $NbS{{e}_{2}}$ electrodes for some time in air.

Because of the relatively low, compared with ordinary metals, electrical conductivity of the starting $NbS{{e}_{2}}$ (${{\rho }_{\parallel }}\sim {{10}^{-4}}\,\Omega \cdot cm$,${{\rho }_{\bot }}\sim {{10}^{-3}}\,\Omega \cdot cm$) the electronic mechanism of heat conduction is relatively weak. In point contacts based on oxidized $NbS{{e}_{2}}$ the thermal conductivity became even lower, and the effects of heating, masking the phonon features, were often very substantial. For this reason, we had to work with contacts characterized by weak oxidation of the $NbS{{e}_{2}}$ surface. As a result, the appearance of "critical" features on the point-contact spectra was not excluded, since superconductivity in the region of the constriction in this case was not adequately suppressed. The latter was confirmed by the comparatively high value of the excess current: right up to values close to theoretical values for clean contacts.

\begin{figure}[]
\includegraphics[width=8cm,angle=0]{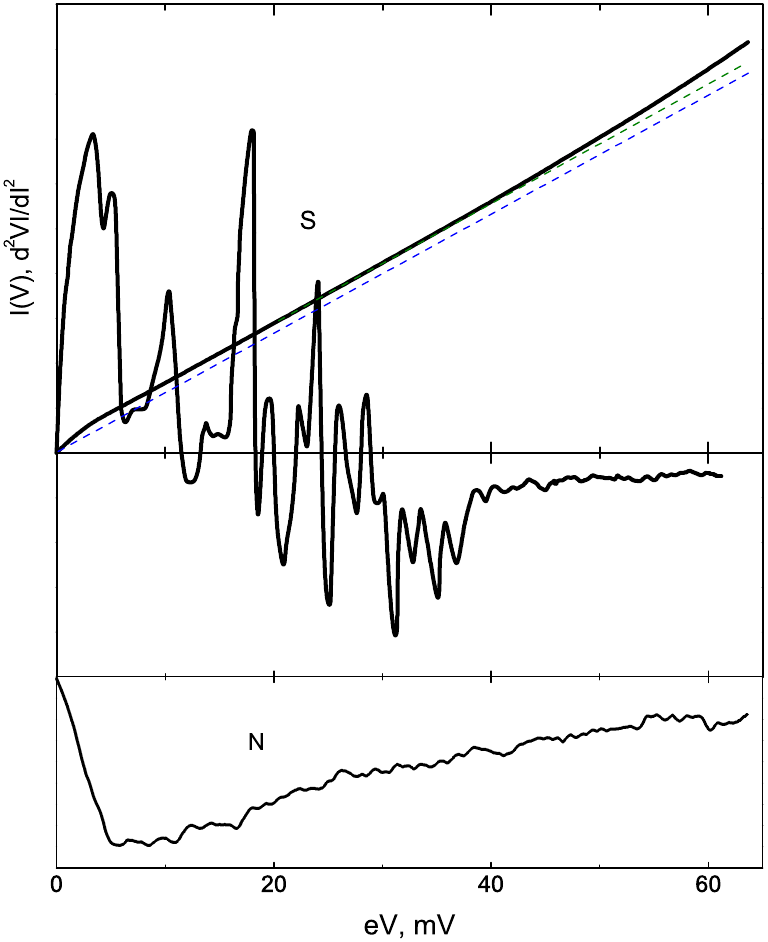}
\caption[]{Current-voltage characteristic and point-contact spectra at a $NbS{{e}_{2}}-Cu$ contacts in the superconducting ($S$) and normal ($N$) states at a temperature of 4.3 and 8$K$, respectively. The axis of the contact is oriented perpendicularly to the $NbS{{e}_{2}}$ layers; $R=90\ \Omega $ and $H=0$.}
\label{Fig1}
\end{figure}

Nevertheless, even in such contacts it was possible to obtain point-contact spectra without "critical" features. An example of such a spectrum for a $NbS{{e}_{2}}-Cu$ contact with an excess current close to the theoretical value for the pure limit is shown in Fig. \ref{Fig1}. The peaks observed in the spectrum presented are bounded by the phonon region of energies and vanish with the transition into the superconducting state, as in the case of dirty $Nb$ contacts \cite{Yanson3}. We note that for the cleanest contacts obtained in our experiments, the excess current could be several times higher than the theoretical value. We have in mind here the theoretical value that is obtained if the value of the energy gap for $NbS{{e}_{2}}$, $\Delta =1.1\,\,meV$ measured by the optical method, is substituted into the formula for the excess current of a clean $S-c-N$ contact \cite{Artemenko}.

For the analysis we chose point-contact spectra without distinct "critical" singularities. The latter had a form approximately corresponding to the narrow peaks in the differential resistance. The width of these peaks, as a rule, did not exceed the characteristic values of the temperature and modulation induced smearing.

A comparative analysis of the point-contact spectra of $NbS{{e}_{2}}$, obtained for the two basic crystallographic directions (parallel and perpendicular to the layers), shows that the phonon peaks characteristic for a given direction are often accompanied by peaks for a second direction also (see Fig. \ref{Fig1}). For measurements along the c axis this is apparently determined by the presence of characteristic growth steps on the lateral surfaces of most $NbS{{e}_{2}}$  crystals. When the shear method for creating point contacts is used, as was done in this work, the latter circumstance greatly facilitated the stratification of this electrode and subsequent tilting of the contact axis away from the given direction. Indeed, when a copper electrode with a sharply ground working face was used, the number of spectra with "secondary" peaks increased markedly. For measurements along the layers, buckling of the plate-like $NbS{{e}_{2}}$  electrodes was entirely possible, even with weak clamping forces because of their small thickness and as a result the orientation of the contact axis could change correspondingly.

\subsection{Point-contact spectra for $2H-NbS{{e}_{2}}$ in the $c$ direction}
\begin{table*}[]
\caption[]{}
\begin{tabular}{|c|c|c|c|c|c|c|c|c|c|c|c|c|c|c|c|c|c|c|} \hline
{Orientation} & \multicolumn{17}{|c|} {Position of features, meV} & Data \\ \hline
\multirow{4}{*}{$\Delta $} & \multicolumn{2}{|c|}{\underbar {\textcolor{red}{5}}}& \underbar{\textcolor{red}{10}} & \multicolumn{2}{|c|}{\textcolor{red}{13}} & \underbar{\textcolor{red}{16}} & \textcolor{red}{19} & \multicolumn{2}{|c|}{\underbar{\textcolor{red}{25}}} & \multicolumn{2}{|c|}{\textcolor{red}{31}} & \multicolumn{2}{|c|}{\underbar{\textcolor{red}{39}}} & \textcolor{red}{44} & \textcolor{red}{48} & \textcolor{red}{52} & \textcolor{red}{58} & This work \\
\cline{2-3}\cline{11-12}
 & 3 & 4 & - & \multicolumn{2}{|c|}{-} & 17 & - & \multicolumn{2}{|c|}{-} & 28 & 30 & \multicolumn{2}{|c|}{-} & - & - & 51 & - & \cite{Feldman} (model 1) \\
 & 3 & 4 & - & \multicolumn{2}{|c|}{-} & 17 & - & \multicolumn{2}{|c|}{-} & 28 & 30 & \multicolumn{2}{|c|}{-} & 42 & - & - & - & \cite{Feldman} (model 2) \\
 & 3 & 4 & - & \multicolumn{2}{|c|}{-} & - & 19 & \multicolumn{2}{|c|}{-} & 29 & 30 & \multicolumn{2}{|c|}{37} & - & - & - & - &\multicolumn{1}{|l|} {\cite{Motizuki}} \\ \hline
\multirow{4}{*}{$\Sigma $} & \multicolumn{2}{|c|}{\underbar{\textcolor{blue}{5}}} & \underbar{\textcolor{blue}{10}} & \multicolumn{2}{|c|}{\textcolor{blue}{14}} & \underbar{\textcolor{blue}{16}} & \underbar{\textcolor{blue}{20}} & \multicolumn{2}{|c|}{\textcolor{blue}{25}} & \multicolumn{2}{|c|}{\underbar{\textcolor{blue}{29}}} & \multicolumn{2}{|c|}{\underbar{\textcolor{blue}{36}}} & \underbar{\textcolor{blue}{42}} & \textcolor{blue}{46-48}&\textcolor{blue}{51-52} & \textcolor{blue}{56-58} & This work \\
\cline{5-5}\cline{6-6}\cline{11-12}
 & \multicolumn{2}{|c|}{-} & 10 & 13 & 14 & 17 & 21 & \multicolumn{2}{|c|}{25} & 29 & 30 & \multicolumn{2}{|c|}{37} & - & - & - & - & \cite{Feldman} (model 1) \\
 \cline{9-10}
 & \multicolumn{2}{|c|}{-} & 10 & 13 & 14 & 17 & 21 & 23 & 26 & 27 & 33 & \multicolumn{2}{|c|}{-} & - & - & - & - & \cite{Feldman} (model 2) \\
  \cline{9-10}\cline{13-14}
 & \multicolumn{2}{|c|}{-} & - & 13 & 14 & 18 & 22 & \multicolumn{2}{|c|}{24} & 28 & 32 & 36 & 37 & - & - & - & - &\multicolumn{1}{|l|} {\cite{Motizuki}} \\ \hline
\end{tabular}
\label{tab1}
\end{table*}

\begin{figure}[]
\includegraphics[width=8cm,angle=0]{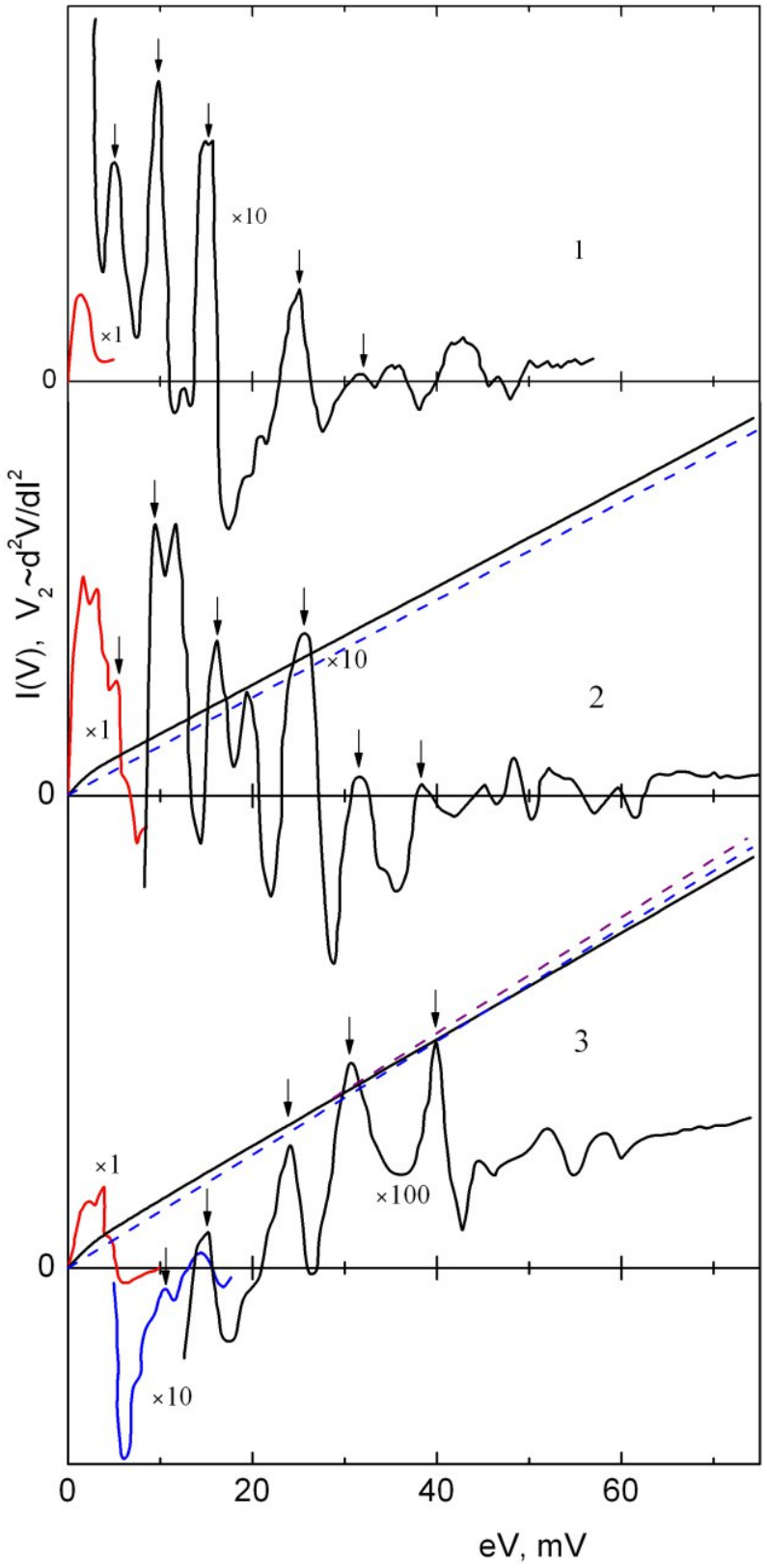}
\caption[]{Current-voltage characteristic and the point-contact spectra of $NbS{{e}_{2}}-Cu$ contacts in the $c-$direction $H=0$: $R$=100 (1), 50 (2), 250 $\Omega $  (3); T= 1.68 (1), 3.94 (2), 3.78(3) $K$. The arrows mark peaks which are best reproduced in different contacts.}
\label{Fig2}
\end{figure}

Figure \ref{Fig2}
\begin{figure}[]
\includegraphics[width=8cm,angle=0]{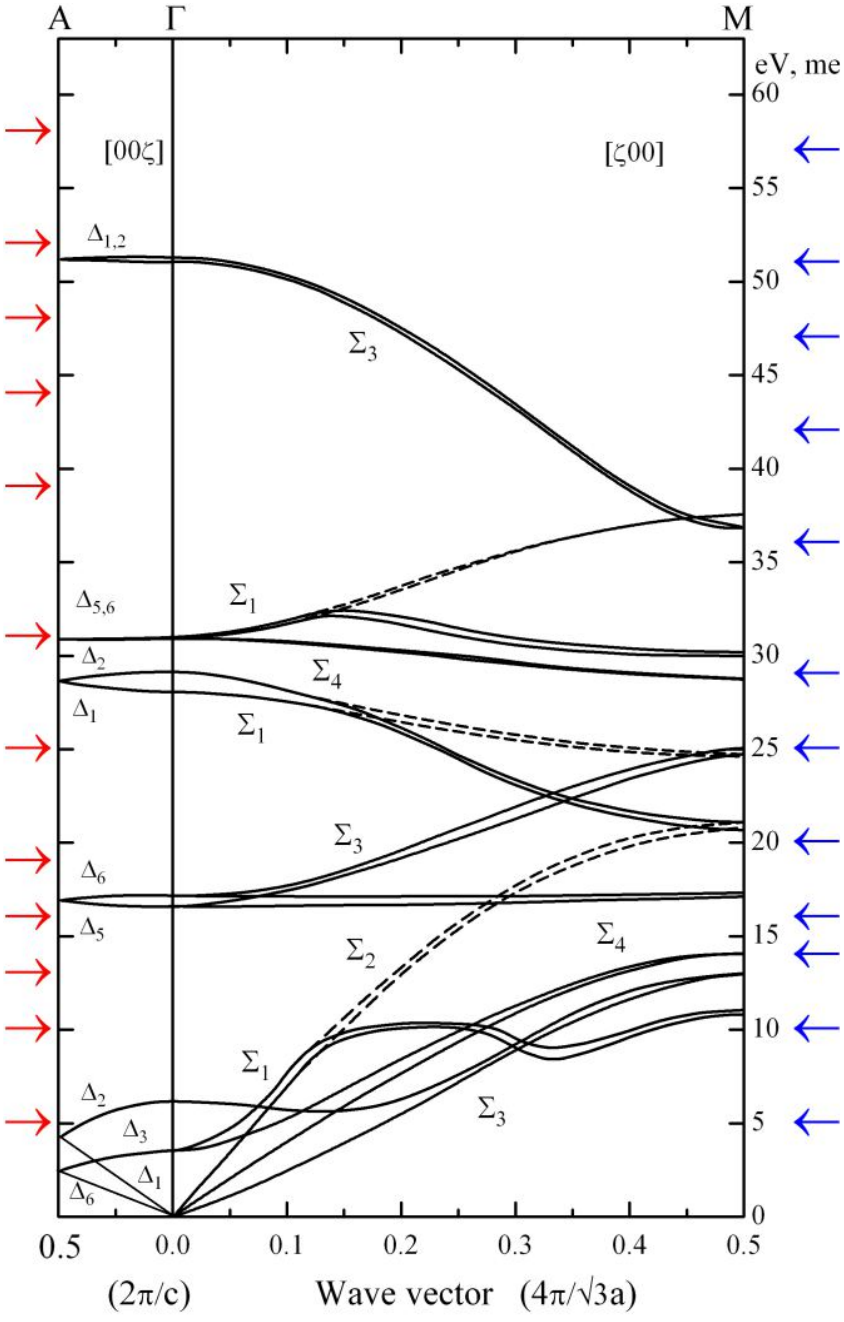}
\caption[]{Computer dispersion curves for $\Delta$ -($\Gamma$$\text{A}$) and  $\Sigma$ -($\Gamma$$\text{M}$) directions of the $2H-NbS{{e}_{2}}$ lattice (model l) \cite{Feldman}. The arrows mark the positions of peaks in the point-contact spectra.}
\label{Fig3}
\end{figure}
shows for three $NbS{{e}_{2}}-Cu$  contacts, whose axes were oriented perpendicular to the $NbS{{e}_{2}}$  layer, the typical point-contact spectra and the I-V curves in the superconducting state, measured in liquid helium at temperatures $T<4.2\text{ }K$. Analysis of the set of point-contact spectra (several tens of spectra), corresponding to a direction perpendicular to the $NbS{{e}_{2}}$ layers, shows that in spite of the substantial resistance of the contact studied (up to several hundreds of ohms), it is often impossible to avoid thermal effects. This is apparently due to the very low thermal conductivity in the $c$ direction. We associate the temperature dependence, observed in a number of spectra, of the position of the phonon peaks (for different contacts the shift by 1-2 $meV$ accompanying a lowering of the temperature by several degrees could occur both toward higher and lower energies) precisely with the heating of the contact region. We note that approximately the same energy spread is observed for the same phonon peaks on different contacts at the same temperature.

The presence of heating effects is also indicated by the appreciable decrease in the excess current and the substantial nonlinearity of the I-V curves, observed for many contacts with large shifts (for example, contact 3 in Fig. \ref{Fig2}). In addition, the possibility of heating of the point-contact region cannot be excluded even in the case of a linear I-V characteristic, since the nonlinearity arising in this case can be compensated by the nonlinearity of opposite sign due to the presence of surface oxide films of a semiconductor character connected in parallel with the micro-short-circuit of the conductivity.

Because of the experimentally observed spread in the energies of the phonon peaks, their position on the energy axis was determined by statistical analysis of 14 spectra, whose character is close to that shown in Fig. \ref{Fig2}. The average values obtained in this case for the positions of the main peaks are presented in Table \ref{tab1} where they are underlined (the error does not exceed 1.5 $meV$). For a number of spectra, weaker (as a rule) additional features were also observed (see Table \ref{tab1}).

We shall compare these data with the theoretical values of the energies of phonons whose wave vectors lie on the boundary of the Brillouin zone and are determined from the computed dependences $\omega (q)$, for which $d\omega /dq\approx 0$  for the $\Delta $  direction \cite{Feldman,Motizuki}. As is evident from Table \ref{tab1} and Fig. \ref{Fig3}, all the main and part of the additional point-contact peaks are comparable, within reasonable limits, with the computed $\omega (q)$ (the best agreement is obtained with the first model from Ref. \cite{Feldman}, if umklapp processes are included and if it is kept in mind that, because of smearing, the computed peaks near 3 and 4 $meV$ may be observed in the experiment in the form of a single wide peak. The possible reasons for the observation of the remaining additional peaks in the point-contact spectra will be discussed below.

\subsection{Point-contact spectra of $2H-NbS{{e}_{2}}$ in the basal surface}

As noted above, in measurements of point-contact spectra in directions parallel to the $2H-NbS{{e}_{2}}$ layers, the orientation of the contact axis in the basal plane was not fixed beforehand. Since the angle between the $\Sigma$ and $T$ directions is equal to ${{30}^{{}^\circ }}$, and the selectivity of the point-contact spectroscopy is limited approximately by the same angles, all peaks characteristic of both directions must appear in the experimental spectra. This should not greatly complicate the observed spectra, since the structures of the vibrational modes for both directions are in many ways similar because of the hexagonal structure of the basal layers \cite{Feldman}.  We have in mind here, of course, the vibrational modes of the starting crystalline lattice (before the transition into the CDW state at 33~$K$).

\begin{figure}[]
\includegraphics[width=8.7cm,angle=0]{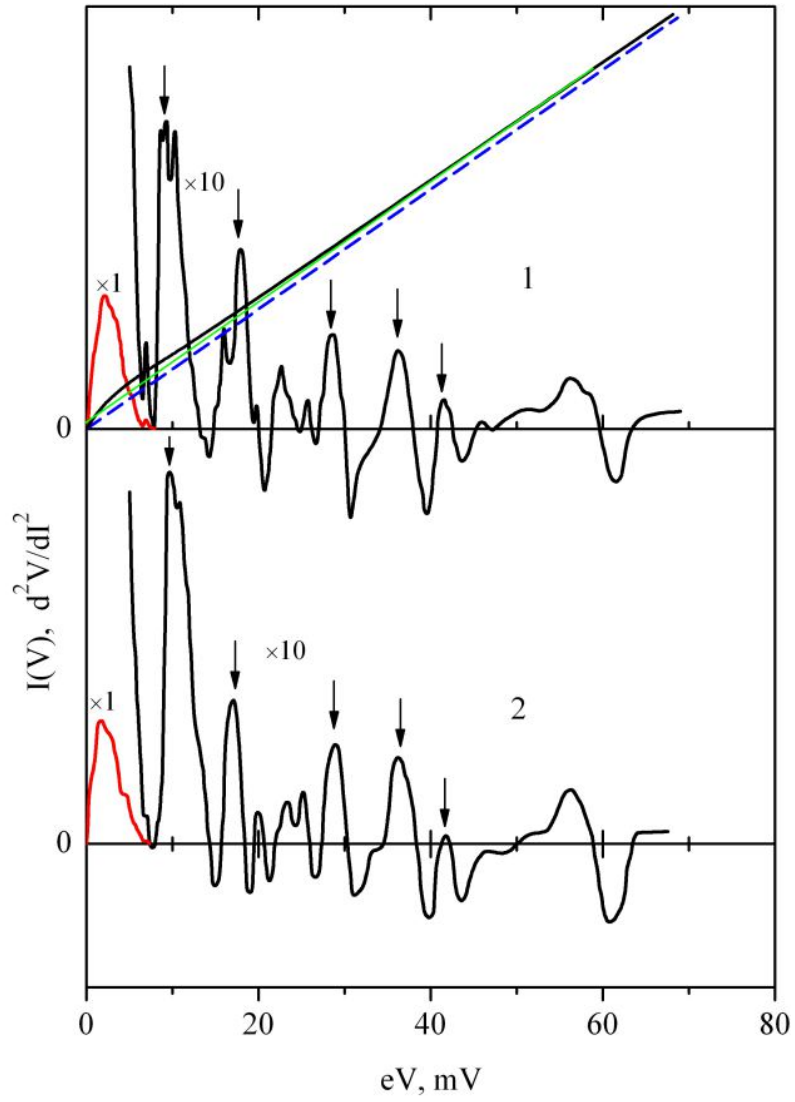}
\caption[]{Current-voltage characteristic and point contact spectra of $NbS{{e}_{2}}-Cu$  contacts in the basal plane for $H$=0 and $T=4.2 K$; $R$ = 1100 and 750 $\Omega $  for the first and second contacts, respectively. Spectrum 2 is for the same contact as spectrum 1, but after spontaneous short-circuit. The arrows indicate the best reproducible peaks for contacts of various groups.}
\label{Fig4}
\end{figure}

The additional phonon modes in the $\Sigma $  direction (the so-called "CDW phonons" \cite{McMillan1}) that appear below the temperature of the phase transition, which are due to the collective vibrations of the amplitude and phase of the CDW, are located primarily in the low-frequency region for the dichalogenides of transition metals. In addition, as follows from experiments on Raman light scattering \cite{Soorvakumar} "CDW phonons" are most distinct only in the commensurate phase. For $NbS{{e}_{2}}$  in the incommensurate phase, the amplitudes mode near 5 $meV$, which has a very high intensity in optical spectra, is an exception. Indeed, for most point-contact spectra in the low-frequency region ($<10\text{ }meV$) only one feature near 5 $meV$ (sometimes in the form of a shoulder on the larger gap peak) was observed, A typical example of this subgroup of spectra is illustrated by two such spectra shown in Fig. \ref{Fig4}. Both spectra refer to the same contact and were obtained as a result of its spontaneous shorting. The positions of the peaks on the energy axis, determined for 10 spectra of this type and reproducible with an accuracy of 1-2 $meV$, are presented in Table \ref{tab1} and underlined. In addition, there are several features that are not as well reproduced. Thus, because of the broadening of the more intense (10 $meV$) peak, the peak near 13-14 $meV$ appears in a number of spectra only as a shoulder on the more intense peak, and the wide peak near 25 $meV$ can split into two peaks. Finally, three peaks are observed in the high-frequency region; in addition, the energy band which they occupy can shift from 46-56 to 48-58 $meV$. The most intense of these three peaks is the peak at the highest energy.

\begin{figure}[]
\includegraphics[width=8.7cm,angle=0]{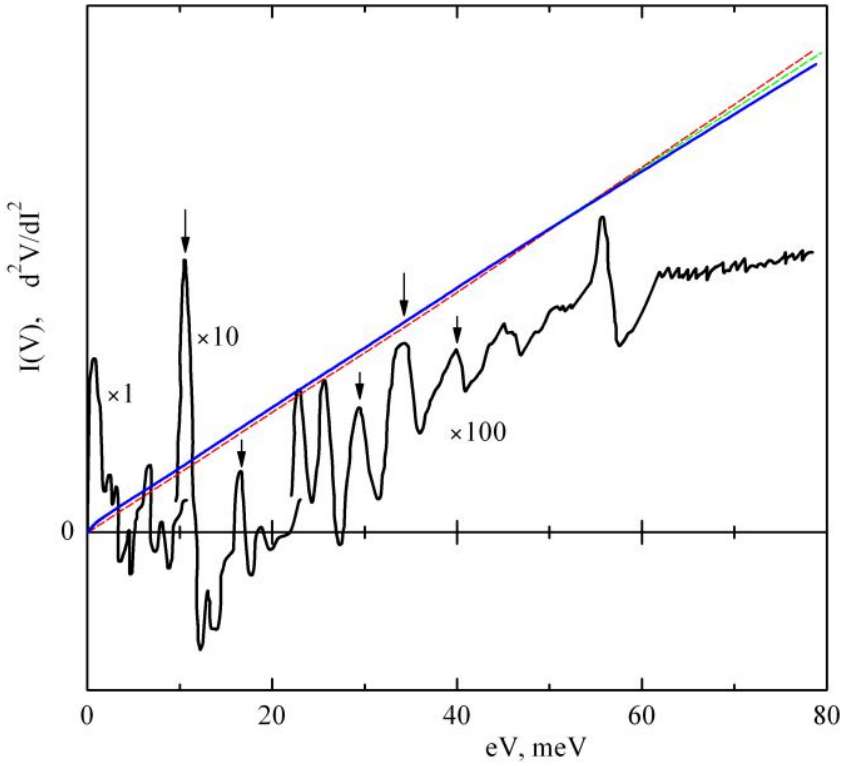}
\caption[]{Current-voltage characteristic and point contact spectrum of a $NbS{{e}_{2}}-NbS{{e}_{2}}$ contact in the basal plane for $H$=0, $R_{0}^{N}=54\ \Omega $, $T=2.79 K$.}
\label{Fig5}
\end{figure}

There exists, however, a second variety of point- contact spectra, in which, aside from the peak at 5 $meV$, other low-frequency ($<10\text{ }meV$) peaks, which are poorly reproduced, were also observed. One of the spectra, measured for the monocontact $NbS{{e}_{2}}-NbS{{e}_{2}}$ , is shown in Fig. \ref{Fig5}. Note the close similarity of the position and form of the main ($eV\ge 10\,\,meV$ ) peaks with the corresponding data obtained for spectra of the first type (see Fig. \ref{Fig4}).

The agreement between the point-contact spectra in the region $eV>10\text{ }meV$, obtained in different series of measurements with random orientation of the contact axis in the basal plane, is in agreement with the proposition, stated in Ref. \cite{Feldman} and confirmed in Ref. \cite{Motizuki}, that the structures of the main lattice modes in the $\Sigma$ and $T$ directions are approximately similar. This enables comparing the positions of the main peaks, observed in the point- contact spectra in directions parallel to the base, with the calculated dispersion curves for the $\Sigma$ direction, presented in Refs. \cite{Feldman} and \cite{Motizuki}.

A comparison shows that the peaks observed in the point-contact spectra in the energy range $10-37 meV$ are in good agreement with the dispersion curves calculated using Feldman's first theoretical model for the dynamic matrix \cite{Feldman}, and the point-contact peak near 5 $meV$ is apparently associated with the amplitude mode of the "CDW phonons." The possible reasons for the appearance of other low-frequency ($<10\text{ }meV$) peaks as well as peaks lying in the high-frequency region ($>37\text{ }meV$) in the experiment will be examined in the next section.

\section{DISCUSSION}
The spectral features that we observed in the second derivative of the I-V characteristic of point contacts consisting of superconducting $NbS{{e}_{2}}$  evidently have a phonon origin, since they appear only in the phonon frequency range (the spectra were recorded up to 100 $meV$) and vanish with the transition from the superconducting into the normal state (see Fig. \ref{Fig1}). This is also confirmed by the fact that most observed peaks in both principal directions (parallel and perpendicular to the layers) can be compared with the corresponding features in the computed dispersion curves.

We emphasize that there is no additional conduction channel for the contacts studied associated with the collective motion of the GDW, a channel that is characteristic of low-dimensional materials below the point of the Peierls transition and experimentally observed in quasi-one-dimensional trichalcogenides of transition metals \cite{Salva}. This could be a result of the effective pinning of CDW on defects by a weak link.

The anisotropy of the electron-phonon interaction, reflecting the anisotropy of the phonon spectrum, is most distinctly manifested in the shift of the high-frequency edge of the EPI spectrum from the range of energies near $\approx 40\ meV$ for contacts whose axis is oriented parallel to the $c$ axis of the $NbS{{e}_{2}}$  crystal (see Figs. \ref{Fig1} and \ref{Fig2}) into the region of energies near 60 $meV$ for contacts oriented parallel to the basal plane (see Figs. \ref{Fig4} and \ref{Fig5}).

The appearance in point-contact spectra of peaks which are not associated with the dispersion curves for the direction fixed by the axis of the contact can be explained by the substantial effect of umklapp processes on the formation of the point-contact spectra of $NbS{{e}_{2}}$, which has a complicated band structure (as done above for the point-contact spectra measured along the $c$ axis). At the same time, the scattering of the electrons with umklapp can cause the wave vector of the phonon, scattering by which reverses the direction of motion of the electrons moving along the $Z$ axis, to be oriented almost perpendicular to the axis of the contact \cite{Yanson1}. This explains the point-contact peaks at 10, 19, 25, and 39 $meV$ for $\Delta $  directions (see Fig. \ref{Fig2}), which do not follow from $\omega (q)$  for this direction, but are characteristic of the $\Sigma$  direction. For the peaks at higher energies, however, such an identification is problematic, since it requires the assumption that some of the dispersion branches for the $\Sigma$  direction are shifted upwards in frequency. Nevertheless, under the assumption that the shift does occur, it is possible to explain the peak at 42 $meV$ in the point-contact spectra in the basal plane, for example, as well as the higher energy features-in the point-contact spectra (up to 48-58 $meV$).

As regards the significant number of additional (with respect to the peak at 5 $meV$) low-frequency peaks ($<~10~meV$) observed in a number of point-contact spectra, the following should be noted here. The appearance of these peaks cannot be ascribed only to the coincidence of the contact axis with the $\Sigma$  direction; otherwise, the peak at 5 $meV$, which for this direction cannot be compared with the lattice branches of the vibrations (see Fig. \ref{Fig3}), would have appeared or disappeared together with the remaining low-frequency peaks. This also follows from the relatively low angular selectivity of point-contact spectroscopy. It is therefore natural to suppose that the only reason that the low-frequency modes appear in the point-contact spectra is that a commensurate section of $NbS{{e}_{2}}$ occurs randomly in the region of the contact, and this would enable electrons to be scattered efficiently by the low-frequency amplitude and phase modes of CDW phonons. (According to McMillan's theory \cite{McMillan2} there should be six such modes, though in experiments on Raman light scattering only four modes were observed in the commensurate phase of $TaS{{e}_{2}}$  \cite{Holy}.) The possibility of the existence of such commensurate sections (domains) in incommensurate phases of chalcogenides of transition metals was predicted theoretically \cite{McMillan1} and confirmed experimentally \cite{Amelinckx}.

Next, if we take into account the strong temperature dependence of the intensity and frequency of optical lines associated with "CDW phonons", \cite{Holy} then we can understand the poor reproducibility of the low-energy peaks in the point-contact spectra measured in the basal plane, since because of the poor thermal conductivity of $NbS{{e}_{2}}$ the temperature in the point-contact region can increase appreciably as the bias is increased. In addition, the reproducibility of these peaks also depends on the ratios of the sizes of the commensurate sections of $NbS{{e}_{2}}$ and the dimensions of the contact.

As regards the reasons for the spread in the energies of the remaining phonon peaks observed in the point- contact spectra, we should note two factors. The first factor is the possible local nonstoichiometry of the chemical composition of $NbS{{e}_{2}}$, right up to the existence of clusters consisting of other phases in this material. Thus in many cases point-contact spectra were recorded, which could not be compared with any of the principal directions of symmetry were recorded, and which could be a result of such phases. The small deviations from stoichiometry of $NbS{{e}_{2}}$ will apparently give rise to smaller basic changes in the point-contact spectra, for example, the small shift in the principal phonon peaks noted above.

Second, as established theoretically, \cite{Khlus} a shift in the phonon peaks in the point-contact spectra toward lower energies by the magnitude of the energy gap relative to the $N$ state is characteristic of clean contacts in the superconducting state. On the other hand, a phenomenological theory of the current-voltage characteristics of the S-c-N microconstrictions, describing the transition from the state of direct conductivity to tunneling, is developed in Ref. \cite{Blonder}. In this case, the excess current vanishes, while the differential conductivity of the contact becomes proportional to the tunneling electron density of states. The latter, as is well known \cite{Giaver,McMillan3}, has features in the region of phonon energies that are determined by the energy dependence of the energy gap $\Delta (E)$, The position of these features on the I-V characteristic and its derivatives is shifted by $\Delta (T)$ toward higher energies relative to the corresponding energies in the phonon spectrum. Thus in the intermediate case, corresponding to the transitional state, the phonon features can be located in the interval $\pm \Delta (T)$  around the corresponding energies of the phonon spectrum; in addition, with the transition from the direct conductivity to tunneling, the form of these features also changes substantially. In addition, the magnitude of the gap for our dirty contacts can apparently vary over a wide range, giving rise to a variable position of the peaks (within the limits of $\Delta$).

An answer has not yet been found to the question of what is responsible for the 1-2 $meV$ temperature shift of the principal ($\ge{10} meV$) phonon peaks. On the basis of the possible overheating of the contact region mentioned above, one reason could be the temperature dependence $\Delta (T)$. Another reason could be linked with the temperature dependence of the relative magnitude of the incommensurateness of the parameters of the main crystal lattice and the CDW lattice, since it is impossible to exclude some effect of the phase transition into the CDW state on the main phonon frequencies in all crystallographic directions.

\section{CONCLUSIONS}
The study of point-contact spectra of superconducting $NbS{{e}_{2}}$ performed in this work enable determining for the first time the approximate position of the main phonon features on the energy axis and establishing the qualitative picture of the behavior of the point-contact EPI function in different crystallographic directions. The low-energy ($<10\text{ }meV$) peaks observed in a number of point-contact spectra are associated with collective oscillations of the amplitude and phase of CDW- the periodically modulated density of conduction electrons, existing in the low-temperature modification of $NbS{{e}_{2}}$ in crystallographic orientations perpendicular to the $c$ axis.

The results obtained confirm the possibility of realizing point-contact spectroscopy of phonons in dirty superconducting contacts \cite{Yanson2}, though the question of the nature of the phonon features in the point-contact spectra in the superconducting state requires further study.

The authors thank I. K. Yanson for formulating the problem and for guiding this work.
\section{NOTATION}
Here $\Delta $  is the energy gap; $\omega $  is the frequency; $g$ is the wave vector; $V$ is the voltage; $I$ is the current; $G$ is the point-contact electron-phonon interaction function; $\rho $  is the resistivity; and $R$ is the resistance of the point-contact.

\end{document}